\documentclass[a4paper,11pt]{article}
\pdfoutput=1 

\usepackage{jcappub} 

\usepackage{graphicx}
\usepackage[mathlines]{lineno}
\usepackage[english]{babel}
\usepackage{color}
\usepackage{ulem}
\usepackage{enumitem}

\def\lapproxeq{\lower .7ex\hbox{$\;\stackrel{\textstyle
<}{\sim}\;$}}
\def\gapproxeq{\lower .7ex\hbox{$\;\stackrel{\textstyle
>}{\sim}\;$}}

\title{Limits on point-like sources of ultra-high-energy neutrinos with the Pierre Auger Observatory}

\author{The Pierre Auger Collaboration}

\affiliation{Observatorio Pierre Auger, Malarg\"{u}e, Argentina}

\emailAdd{auger\_spokespersons@fnal.gov}

\abstract{With the Surface Detector array (SD) of the Pierre Auger Observatory we can detect neutrinos with energy between $10^{17}\,$eV and $10^{20}\,$eV from point-like sources across the sky, from close to the Southern Celestial Pole up to $60^\circ$ in declination, with peak sensitivities at declinations around $\sim -53^\circ$ and $\sim+55^\circ$, and an unmatched sensitivity for arrival directions in the Northern hemisphere. A search has been performed for highly-inclined air showers induced by neutrinos of all flavours with no candidate events found in data taken between 1 Jan 2004 and 31 Aug 2018. Upper limits on the neutrino flux from point-like steady sources have been derived as a function of source declination. An unrivaled sensitivity is achieved in searches for transient sources with emission lasting over an hour or less, if they occur within the field of view corresponding to the zenith angle range between $60^\circ$ and $~95^\circ$ where the SD of the Pierre Auger Observatory is most sensitive to neutrinos.}


\begin{document}

\maketitle
\flushbottom

\def\linenumberfont{\normalfont\tiny\itshape\sffamily}

\section{Introduction}

There are inherent difficulties in establishing the acceleration mechanisms and the sources of Ultra-High-Energy Cosmic Rays (UHECRs). This is in great part due to the bending of the paths of the charged particles in the intervening magnetic fields, because neither the particle charge, determined by the primary composition, nor the amplitude and direction of the magnetic fields are known.  Recent results on anisotropies in the arrival directions of UHECRs~\cite{Dipole_Science_2017,Dipole_ApJ_2018,Auger_SBG_2018} provide valuable but still limited information. Efforts to combine the spectral shape and the primary composition~\cite{Auger_combined_fit_JCAP_2017} and to search for diffuse fluxes of Ultra-High-Energy (UHE) photons \cite{Auger_hybrid_photons_JCAP2017} and neutrinos~\cite{IceCube_PRD2018,Auger_nus_PRD2015,Auger_nus_PRD2019} have proven useful in constraining production models, but the results are not yet conclusive. 
Astrophysical neutrinos in the UHE band, exceeding $\sim\,1$~EeV ($10^{18}$~eV), are naturally expected in association with UHECRs. UHE neutrinos must be produced as a result of collisions of UHECRs with matter and radiation within the sources that accelerate them~\cite{Becker_PhysRep_2008}, during transport to earth~\cite{BZ_cosmogenic_1969,Engel_GZK_2001}, or both. 
Not surprisingly, measuring the neutrino flux is one of the highest priorities  because of the directional information they carry. The measurement of the diffuse flux of astrophysical neutrinos in the $10^{14}$~eV to $10^{15}$~eV range~\cite{Aartsen:2015knd} was a great milestone in astroparticle physics. However, no clear identification of the sources has yet been obtained from the arrival directions of the neutrinos alone. 

The birth of gravitational wave (GW) astronomy~\cite{LIGO_GW150914,LIGO_GW151226} has given an impressive boost to multimessenger astronomy, particularly with the recent detection of the merger of two neutron stars~\cite{GW170817_BNS}. This landmark detection has triggered observations in practically all bands of the electromagnetic spectrum, including setting stringent limits on the high-energy neutrino flux with current neutrino telescopes that constrain several models of neutrino production in these objects~\cite{GW170817_BNS_nus}.
Also, a correlation between a $\sim 300$ TeV neutrino and a powerful blazar, TXS 0506+056, from the same position in the sky has recently been reported~\cite{IceCube_TXS}. The blazar was also observed in an active state
with gamma-ray emission between 50 GeV and a few hundred GeV detected at the time of the neutrino observation~\cite{MAGIC_TXS}. Moreover, a search for neutrinos from this direction revealed a burst of 13 neutrinos detected with IceCube during a period of 156 days between December 2014 and February 2015~\cite{IceCube_TXS_prior}. While this correlation suggests blazars as a potential site for PeV neutrino production, the picture is far from complete, and it is also possible that there are contributions from other sources~\cite{Padovani_TXS}. Naturally, there is much interest in detecting neutrinos at energies beyond those already detected by IceCube, all the way to the UHE regime, to help unveil the origin of the detected PeV neutrinos, of the UHECRs and the possible connections between them \cite{Astro2020_White_Paper}. 

The Pierre Auger Observatory, an array of particle detectors~\cite{Auger_SD_2008,Auger_observatory_NIMA2015} and several fluorescence telescopes~\cite{Auger_FD_2010,Auger_observatory_NIMA2015} located in Malarg\"ue, Argentina, is the largest and most precise detector for air showers induced by UHE particles. The Observatory has been running since 2004, well before it reached its final design size of 3000~km$^2$ in 2008. Its design also considered the search for UHE neutrinos by  looking for inclined showers that develop deep in the atmosphere~\cite{Capelle_1998}, and it was later shown that it is particularly effective for Earth-Skimming (ES) tau neutrinos. These ES neutrinos convert below the surface to tau leptons that exit to the atmosphere and decay, inducing an air shower~\cite{ES_tau_neutrinos}. Methods have been devised to identify the showers produced by neutrinos in the background of cosmic-ray showers. The data have been scanned for fluxes of UHE neutrinos, and no candidates have been found~\cite{Auger_nus_PRD2019}. These searches have led to limits on both the diffuse flux~\cite{Auger_nus_PRD2015,Auger_nu_ICRC17} and point  source fluxes~\cite{Auger_point_like_ApJ2012,Auger_nu_ICRC17} of UHE neutrinos. 

In this paper, we describe the directional sensitivity and sky coverage of the Surface Detector Array of the Pierre Auger Observatory to UHE neutrinos. 
We give upper limits to the neutrino flux as a function of equatorial declination after the analysis of data between 1 Jan 2004 and 31 Aug 2018, updating the previously published limits that were obtained with a reduced data set between 1 Jan 2004 and 31 May 2010~\cite{Auger_point_like_ApJ2012}. Specific details on the point source search for GW170817 and TXS 0506+056 are reported separately, in \cite{GW170817_BNS_nus} and \cite{TXSInPreparation}, respectively.

\section{The neutrino search at the Pierre Auger Observatory}
\label{sec:selection}

The Pierre Auger Observatory combines an array of 1660 particle detectors spread over an area of $\sim\,3000$~km$^2$ in a hexagonal pattern, the Surface Detector Array (SD), each separated 1.5 km from its six nearest neighbours, and 27 telescopes, the Fluorescence Detector (FD), that view most of the atmosphere over the SD~\cite{Auger_observatory_NIMA2015}. The particle detectors, each filled with 12 tons of purified water, register the Cherenkov light emitted when charged particles from the shower front go through, while the FD telescopes capture the fluorescence light emitted by the nitrogen of the atmosphere when excited by the passage of the shower front. The signals in the particle detector stations are digitized in 25 ns bins and temporarily stored when they satisfy the second level trigger (T2): either a threshold for the peak of the signal or a smaller threshold in at least 13 bins, the ``Time-over-Threshold" (ToT) trigger. The resulting signal traces of at least three T2-stations correlated in space and time are stored as long as the event passes a third level trigger (T3), designed to accept showers and reject accidental coincidences. Cosmic-ray showers are detected with increasing efficiency as the shower energy rises~\cite{Auger_trigger_eff_NIMA2010}, reaching practically $100~\%$ above 4~EeV for zenith angles as high as $\theta=80^\circ$~\cite{Auger_spectrum_HAS_JCAP15}. 
More details can be found in~\cite{Auger_trigger_eff_NIMA2010}.  

The identification of showers induced by neutrinos is easily achieved by looking for inclined showers that develop close to the ground. Cosmic-ray particles arriving at a large inclination with respect to the vertical interact in the upper parts of the atmosphere and must traverse much larger depths than vertical showers to reach the ground. When the shower front reaches the ground level, the shower has developed well beyond its maximum number of particles and the electromagnetic component has been practically absorbed. The front contains mostly 20--200~GeV muons that have hardly undergone any interaction, except for continuous energy loss and deviations in the magnetic field of the earth. These muons accumulate smaller delays than electrons and photons with respect to an imaginary particle moving along the shower axis at the speed of light~\cite{Cazon:2003ar}, and leave characteristic faster and shorter pulses in the signal traces recorded by the SD stations. 

The basic strategy to search for UHE neutrinos in the Pierre Auger Observatory consists of selecting inclined showers and ensuring that they have a large electromagnetic component at ground level compared to the cosmic-ray background. Naturally, this depends on the zenith angle and, for this reason, the selection has been split into two channels: Earth-Skimming~(ES) neutrinos that for this observatory have been shown to concentrate in a narrow range of zenith angles between $\theta=90^\circ$ and  $\theta=95^\circ$ at EeV energies~\cite{Zas_NJP2005,AugerNuTau2008}, and downward-going (DG) with zenith angles between $\theta=60^\circ$ and $\theta=90^\circ$. The conversion mechanisms of a neutrino into an air shower are different for DG and ES, and the requirements made on the signals to efficiently separate neutrinos from background events also call for different strategies. For optimization purposes, the DG procedure is further subdivided into two sets for Low zenith angles (DGL) ~\cite{Auger_DGL,Auger_nus_PRD2015}, between $\theta=60^\circ$ and $\theta=75^\circ$, and High zenith angles (DGH)~\cite{Auger_DGH_PRD2011,Auger_nus_PRD2015}, between $\theta=75^\circ$ and  $\theta=90^\circ$. 

The selection in each of these categories is made using a different set of parameters adapted to the different zenith-angle ranges of each channel. 
An important difference is that the search is performed in all triggered events for ES showers while only events with four or more triggered stations are considered for the DG events to reduce the background. The angular selection of ES is based on a high eccentricity of the elongated signal patterns on the ground, on the average of the speed along the major axis of the ellipse with which the signal appears to move on the ground between pairs of stations, which must be the speed of light $c$ with a small tolerance ($\sim\,4$\%), and on its RMS value ($<0.26c$)~\cite{AugerNuTau2008,Auger_nus_PRD2015}. 
For DGH events, the angular selection is made requiring also a high eccentricity of the signal pattern, an average apparent speed close to $c$ (with a similar tolerance) and a small RMS value ($<0.08c$). The reconstructed zenith angle of the shower, assuming a plane shower front, is required to be greater than $75^\circ$. Finally, for DGL events the selection is based on reconstructed zenith angles between $58.5^\circ$ and $76.5^\circ$ to allow for reconstruction uncertainties, and a requirement that at least $75\%$ of the stations have a ToT trigger to select signals spread in time (see \cite{Auger_nus_PRD2015,Auger_nus_PRD2019} for further details). The arrival direction reconstruction has not been optimized for neutrino events, and it is just used for down-going events as one of the multiple strategies to select inclined showers. Its effectiveness when combined with all the other selection criteria has been studied using simulations of neutrino-induced showers.

In each channel, the selection of the neutrino-induced showers is performed making a cut on a single parameter which is related to the width of the signal traces of the triggered detectors. The choice of parameter and the cut have been optimized by comparing extensive  simulations of neutrino events to a small fraction of the data, assumed to be cosmic-ray background. The cut is chosen at a value of the parameter such that only one event in each selection could be expected after a number of years of observation from the extrapolation of the background distribution (see \cite{Auger_nus_PRD2019} for details). The parameter is different for each case~\cite{Auger_nus_PRD2015,Auger_nus_PRD2019}. In the ES channel, the average Area over Peak ($\langle{\rm AoP}\rangle$), defined as the average over stations of the ratio of integrated charge of the trace and its maximum value, has proved a good choice. The AoP of an individual station is normalized to one for single muons used for SD station calibration~\cite{Auger_observatory_NIMA2015}. A cut at $\langle{\rm AoP}\rangle =1.83$ selects neutrinos very efficiently (about $95\%$ of those that trigger the SD). For the DG cases, a multivariate Fisher discriminant method has been used combining the AoP of selected stations. The DGH case combines nine variables obtained from the AoP of the four earliest stations and a tenth measuring the asymmetry in AoP between the earliest and latest stations. The values of the cuts are optimized separately for three sub-sets within the DGH channel depending on the number of triggered stations. In the DGL case, a Fisher discriminant based on five or six parameters obtained from the 4 or 5 stations closest to the shower core is constructed. Five sub-groups have been made in this channel subdividing the zenith angle in five bands, and the cut on the Fisher discriminant is calculated separately in each of them. More details can be found in~\cite{Auger_nus_PRD2015, Auger_nus_PRD2019}.
As a result of this optimization procedure, the effective areas addressed in the following section display some discontinuities at $\theta=75^\circ$ and $90^\circ$, the limiting zenith angles 
between the three different search procedures.

\section{Sensitivity of the Observatory to point-like neutrino sources} 

Each neutrino search category, ES, DGH and DGL, corresponds to a given range of zenith angles, and the three categories combined cover the range between $\theta=60^\circ$ and $\theta=95^\circ$. Naturally, the neutrino identification efficiency is different in each category, making the sensitivity of the Observatory dependent on the direction in the sky where the search is performed. 

\subsection{Effective Area} 

\begin{figure}[!tb]
\centering
\includegraphics[width=0.9\textwidth]{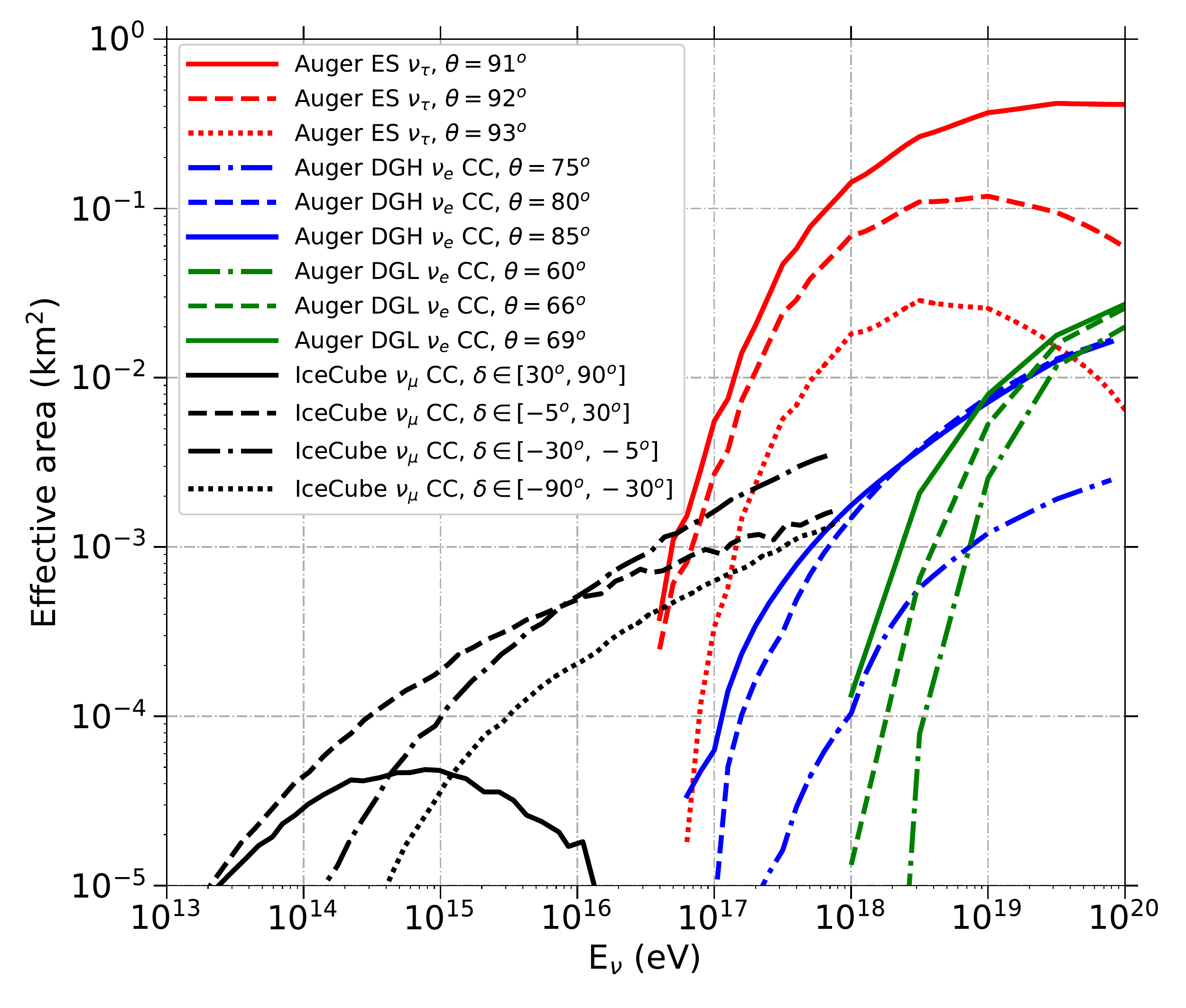}
\caption{Instantaneous effective areas for ES (red lines), DGH (blue) and DGL (green) channels as a function of neutrino energy for selected zenith angles ($\theta$) as labelled, compared to those of IceCube~\cite{IceCube_PS_2017} (black). The DG and ES effective areas are respectively obtained with Eqs.~(\ref{eq:EffectiveAreaDG}) and (\ref{eq:EffectiveArea_ES}).
Numerical values of the effective areas for ES, DGH and DGL analyses as a function of energy and zenith 
angle are available at \cite{link_effective_areas}.
For IceCube at a latitude $\lambda=-90^\circ$, the zenith angle $\theta$ and the declination 
$\delta$ are related by $\theta=90^\circ+\delta$.
}
\label{fig:effective_area}
\end{figure}

\begin{figure}[!tb]
\centering\includegraphics[width=0.9\textwidth]{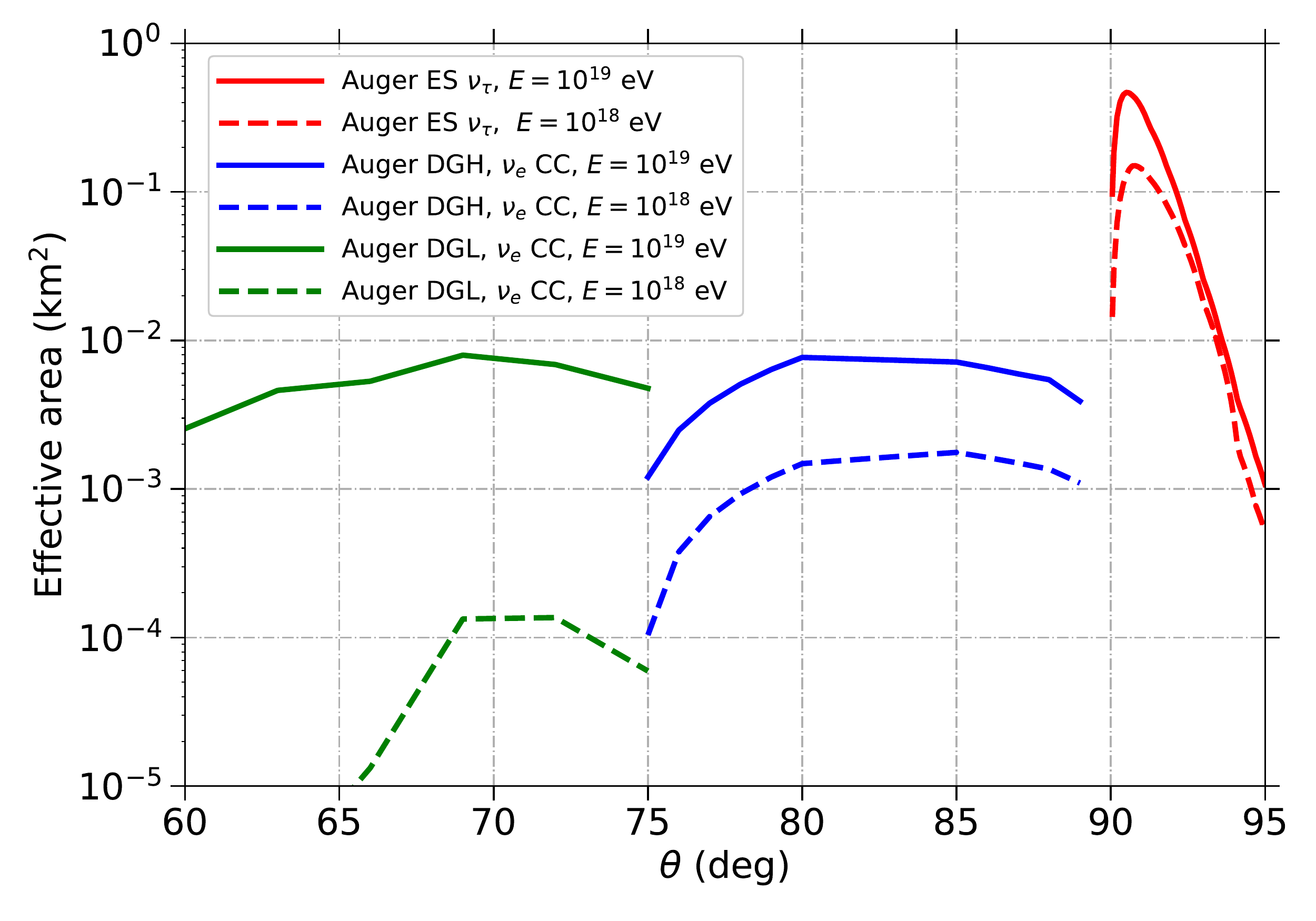}
\caption{
Instantaneous effective areas for ES (red lines), DGH (blue) and DGL (green) neutrinos as a function of zenith angle for selected neutrino energies. The DG and ES effective areas are respectively obtained with Eqs.~(\ref{eq:EffectiveAreaDG}) and (\ref{eq:EffectiveArea_ES}).}
\label{fig:effective_areaVsTheta}
\end{figure}

The sensitivity in each direction can be quantified in terms of the effective area ${\cal A}_i(E_\nu)$, to neutrinos of flavour $i=\nu_e,~\nu_\mu,~\nu_\tau$ and energy $E_\nu$, defined such that ${\cal A}_i$ multiplied by the spectral flux of flavour $i$ from a point source, $\phi_i(E_\nu)={\rm d}^4N/({\rm d}E_\nu~{\rm d}A~{\rm d}t)$, gives the energy spectrum of the instantaneous rate of detected events. The rate of detected events is obtained by integrating it over energy:
\begin{equation}
{{\rm d} N_i \over {\rm d}t} = \int_{E_\nu}~{\rm d}E_\nu ~ \phi_i(E_\nu)~{\cal A}_i(E_\nu). 
\label{eq:EventRate}
\end{equation}
Each neutrino flavour must be treated separately because the showers they initiate through charged-current (CC) interactions are substantially different in the fraction of energy that they carry relative to the incident neutrino~\cite{Auger_DGH_PRD2011,Auger_DGL}. For DG showers, the effective area is obtained by integrating the neutrino identification efficiency, $\varepsilon_{i,c}$, and the interaction probability per unit depth\footnote{This is just the inverse of the energy-dependent neutrino-nucleon mean free path in g~cm$^{-2}$.} $\sigma_c~m_p^{-1}$, where $m_p$ is the mass of a proton, and $\sigma_c$ the neutrino-nucleon cross-section, over the array area $A$, (transverse to the neutrino direction) and over the atmospheric matter depth of the neutrino trajectory $X$:
\begin{equation}
{\cal A}^{\rm DG}_{i,c}= \int_X\!\int_A\! {\rm d}X~{\rm d}A~\cos\theta ~ \varepsilon_{i,c}~ \sigma_c~m_p^{-1}.
\label{eq:EffectiveAreaDG}
\end{equation}
Both $\varepsilon_{i,c}$, and $\sigma_c$ are different for neutral- (NC) and charged-current (CC) interactions (index $c$), but at the energies of interest, the change in the cross-sections for different flavours is negligible. The efficiency is calculated using simulations of extensive air showers. It includes an average over possible momentum fractions transferred to the nucleus in the collision, and it depends strongly on both energy and zenith angle of the neutrino. As a function of neutrino interaction depth $X$, the efficiency is maximized when $X$ is such that the shower maximum is approximately reached at ground level. 
In addition, it also depends on the impact point of the shower at the ground and on the instantaneous configuration of the SD (which changed substantially until deployment was completed in May 2008). There are thus instantaneous effective areas ${\cal A}_{i,c}^{\rm DG}$ for DG showers for each flavour $i$ and interaction type $c$. Naturally, the effective areas also depend strongly on neutrino energy and on zenith angle (see Figs.\ \ref{fig:effective_area} and \ref{fig:effective_areaVsTheta}). 

For the ES case, the calculation of the effective area is much more involved. The differential probability that a tau neutrino of energy $E_\nu$ undergoes a CC interaction along the earth's chord, and that the resulting tau lepton exits to the atmosphere with energy $E_\tau$, denoted as $p_{\rm exit}(E_\nu,E_\tau,\theta)$, is given by a similar integral of the interaction probability per unit depth along the neutrino trajectory. $p_{\rm exit}$ is obtained with a dedicated MC simulation that also takes into account neutrino absorption along the earth's chord, regeneration of the neutrino flux both through NC interactions and tau decays underground, energy loss of the tau lepton while it travels through the earth and the survival probability to reach the earth's surface~\cite{Payet_PRD2008,NuTauSim_2018}. The rapid rise of the earth's chord as the zenith angle increases below the horizon and its absorptive effect for high-energy neutrinos 
are responsible for a strong dependence of $p_{\rm exit}$ on $\theta$ within a small range from $\theta=90^\circ$ to $\theta \simeq 95^\circ$ \cite{Bertou_APP2012,Zas_NJP2005,Payet_PRD2008,NuTauSim_2018}. This probability, folded with the selection and identification efficiency $\varepsilon_{\rm ES}$ and the tau decay probability per unit length, must be integrated over tau energy and decay length $l$ to obtain the effective area:    
\begin{eqnarray}
{\cal A}^{\rm ES} = 
\int_{E_\tau}\!\int_A\!\int_l\,{\rm d}A~{\rm d}E_\tau~ 
{{\rm d}l~\over \gamma_\tau\lambda} \exp\left[-{l \over \gamma_\tau\lambda}\right] 
\vert\cos\theta\vert ~p_{\rm exit}~\varepsilon_{\rm ES}, \nonumber \\
\label{eq:EffectiveArea_ES}
\end{eqnarray}
where $\lambda=c\beta_\tau\tau_\tau \simeq86.93\times10^{-6}$~m is the decay length, $\beta_\tau$ and $\gamma_\tau=E_\tau/(m_\tau c^2)$ are the speed and Lorentz factor of the tau lepton, $m_\tau\simeq 1.777$~GeV is its mass, and the tau-lepton is assumed to be ultra-relativistic.  

The instantaneous effective area for the ES, DGH and DGL neutrinos as a function of neutrino energy is displayed in Fig.~\ref{fig:effective_area} for selected zenith angles and is compared to that of IceCube~\cite{IceCube_PS_2017}. The EeV energy range in which the Pierre Auger Observatory has optimal effective area extends in energy beyond the published effective area of IceCube and, for favourable source positions as seen from the SD, the effective area of the Pierre Auger Observatory is significantly larger.

The dependence of the effective area on the zenith angle is displayed in  Fig.~\ref{fig:effective_areaVsTheta}, for DG charged-current electron neutrinos and selected neutrino energies in the zenith angle range from $\theta=60^\circ$ to $\theta=90^\circ$, and for ES events from $\theta=90^\circ$ to $\theta=95^\circ$. A strong dependence on $\theta$ can be clearly seen in the ES range and at $\theta=90^\circ$, the transition from DG to ES. At $10^{18}$~eV the ES effective area at $\theta=91^\circ$ is nearly two orders of magnitude higher than for DG electron neutrinos with CC interactions at $\theta\sim 80^\circ$. The latter is the channel giving the largest contribution to the DG calculation because the probability of CC interactions is $\sim 2.5$ times greater than neutral currents, and moreover, in the case of the electron flavour, all the neutrino energy gets transferred to the induced shower. 
The ES search is notably more sensitive than the DG searches due to several reasons. The matter depth for neutrino interactions along the earth's chord is much larger than the available depth of the atmosphere for DG showers and, at zenith angles very close to the horizontal, the conversion probability is maximized for energies just below the EeV~\cite{Bertou_APP2012,Zas_NJP2005,Payet_PRD2008,NuTauSim_2018}. In addition, the ES search considers events with at least three stations making this channel more efficient to detect lower-energy showers than in DG. 

\subsection{Sky Coverage} 

\begin{figure}[!tb]
\centering
\includegraphics[width=0.9\textwidth]{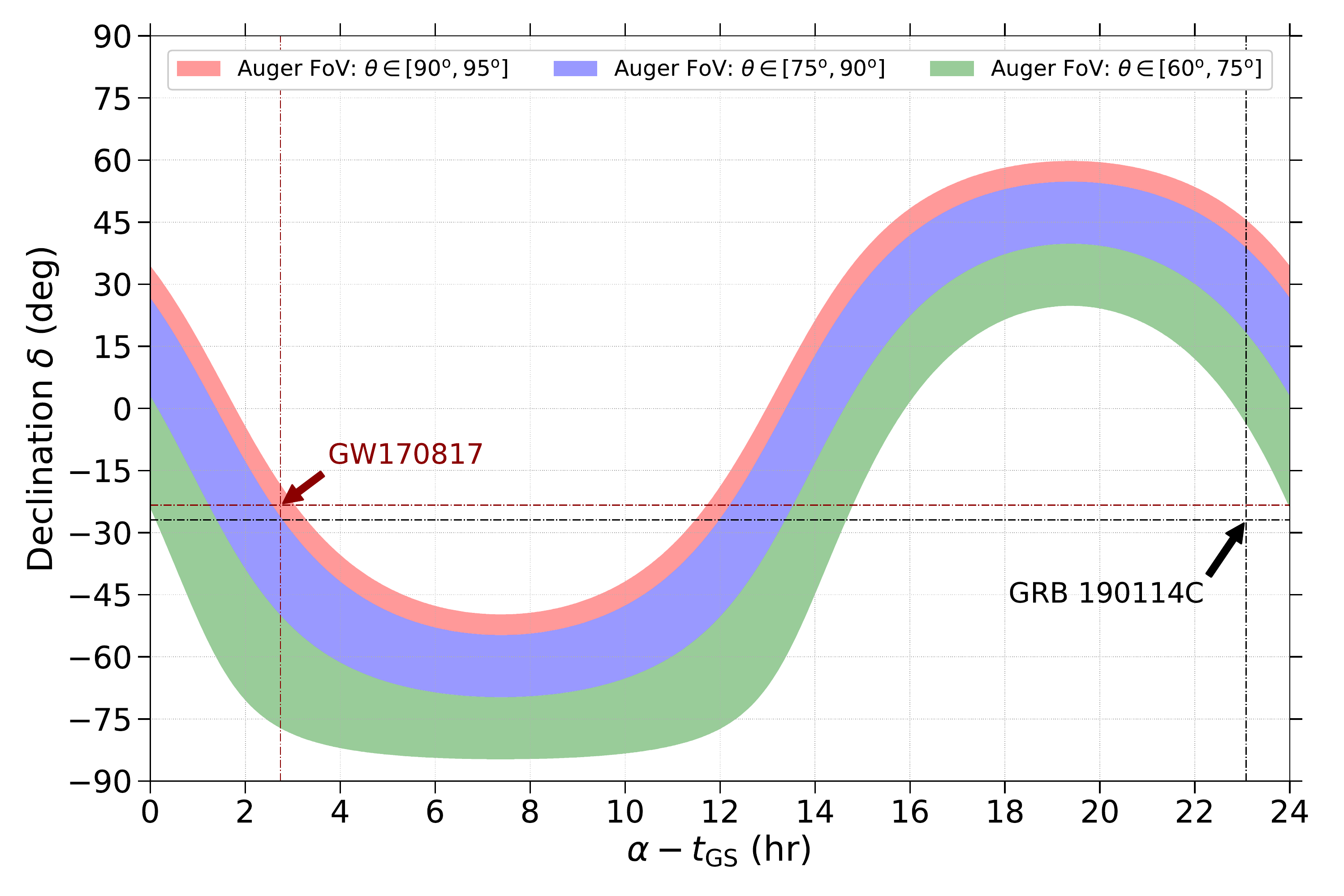}
\caption{Instantaneous field of view (FoV) of the Pierre Auger Observatory for ES, DGH and DGL neutrinos as a function of the hour angle. The declination range for a given location in right ascension, $\alpha$, at a given Greenwich Sidereal Time (converted to angle), $t_{\rm GS}$, is directly read at the corresponding hour angle: $\alpha-t_{\rm GS}$ (see text). Two source examples are shown: GW170817 visible in Auger in the ES at the time of emission \cite{GW170817_BNS_nus}, and GRB 190114C \cite{GRB190114C} not visible in Auger in the inclined directions at the time of the burst.}
\label{fig:FoVDecHourAngle}
\end{figure}

As the neutrino search at the Pierre Auger Observatory is limited to 
showers with $\theta$ between $90^\circ$ and $95^\circ$ in the ES analysis, and between $60^\circ$ and $90^\circ$ in the DG analysis, at each instant, neutrinos can be effectively detected only from a specific region of
the sky corresponding to this range. A point-like source at a declination $\delta$, right ascension $\alpha$ (equatorial coordinates) and a local sidereal time $t$ is seen from the latitude of the Observatory ($\lambda = -35.2^\circ$) with a time-dependent zenith angle $\theta(t)$ given by:
\begin{equation}
\cos\theta(t) = \sin\lambda\,\sin\delta +
\cos\lambda\,\cos\delta\,\sin\left(2 \pi \frac{t}{T} - \alpha\right),
\label{eq:costheta-t}
\end{equation}
where $T$ is the duration of one sidereal day, and the angle in brackets is the  so-called hour-angle. At any given instant, the field of view (FoV) of the Observatory for neutrino search is limited by the imposed restrictions on $\theta$. In fact, the three searches ES, DGH and DGL correspond to different fields of view. This can be seen in  Fig.~\ref{fig:FoVDecHourAngle}, where the corresponding FoV bands are plotted in equatorial coordinates as a function of $\alpha-t_{\rm GS}$, where $t_{\rm GS}=2\pi~t/T+\ell$ is the Greenwich Sidereal Time (GST) converted to angle and $\ell$ is the mean longitude of the Observatory. For any given $\alpha$, the instantaneous declination range for neutrino search at 00:00 GST ($t_{\rm GS}=0$) can be directly read from the plot at a value $\alpha$ of the abscissa. At any other $t_{\rm GS}$, the corresponding declination range is simply read from  Fig.~\ref{fig:FoVDecHourAngle} at $\alpha - t_{\rm GS}$. 

As can be seen in Fig.~\ref{fig:FoVDecHourAngle}, the SD of the Pierre Auger Observatory is sensitive to point-like sources of neutrinos over a broad declination range between $\delta \sim -85^\circ$ and $\delta \sim 60^\circ$. The search for ES showers is limited to a narrower band between $\delta \sim -55^\circ$ and $\delta \sim 60^\circ$ (red band), for DGH showers between $\delta \sim -70^\circ$ and $\delta \sim 55^\circ$ (blue band), and for DGL showers between $\delta \sim -85^\circ$ and $\delta \sim 40^\circ$ (green band) - see also Fig.~\ref{fig:PS-Fractime}. 

Eq.~(\ref{eq:costheta-t}) also illustrates that the detector location has a large impact on the sensitivity to point-like sources of neutrinos. In particular, for a detector such as IceCube located at the South Pole at a latitude $\lambda=-90^\circ$, Eq.~(\ref{eq:costheta-t}) is reduced to $\cos\theta=-\sin\delta$, and as a consequence, a source at a given declination is always seen with the same zenith angle. This is not the case for the Pierre Auger Observatory, where a given point in the sky has a zenith angle that is varying with a period of one sidereal day. The point source sensitivity of the Pierre Auger Observatory is, therefore, a direct result of the long-term averaging of its periodically changing field of view.

\subsection{Exposure}
\label{subsec:exposure}

\begin{figure*}[!htb]
\centering
\includegraphics[width=0.9\textwidth]{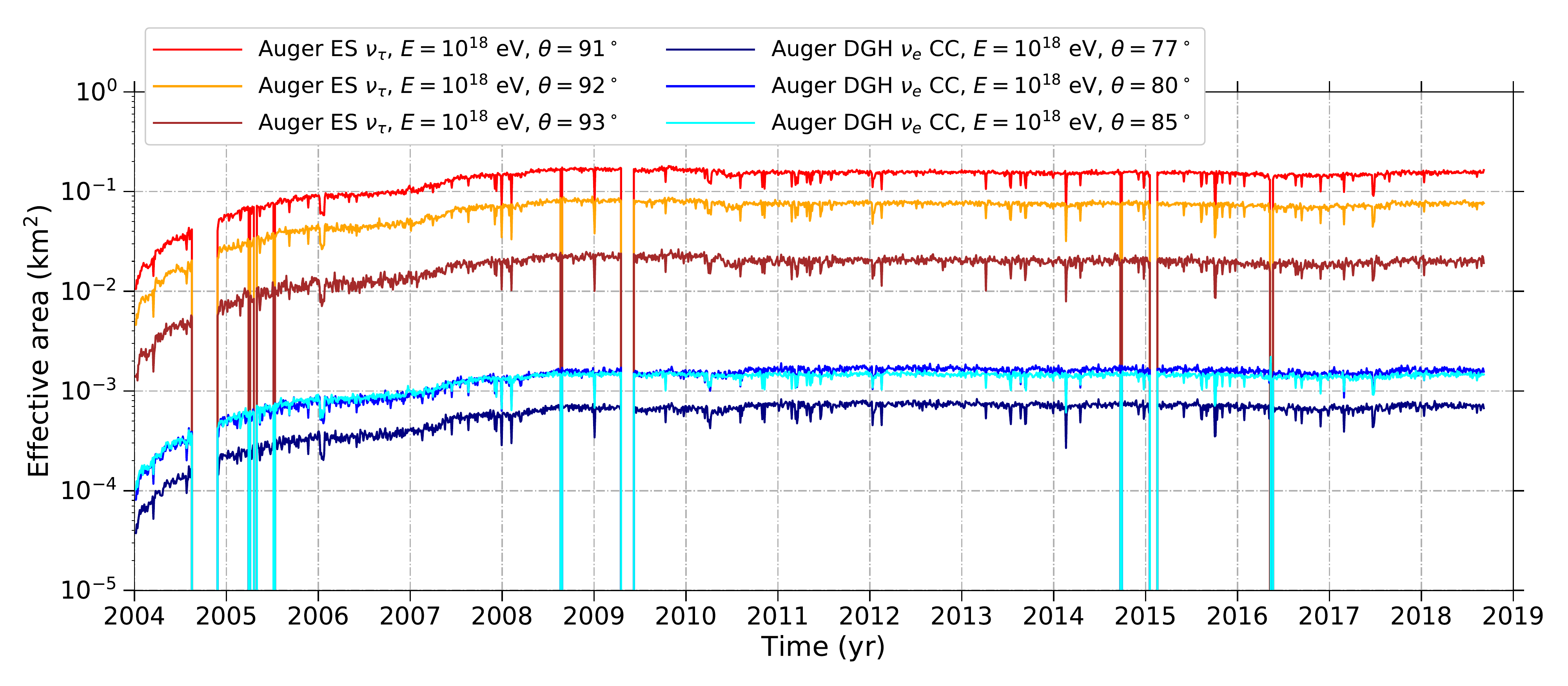}
\caption{Average effective area calculated for three-day intervals for the ES and DGH channels as a function of time for the data taking period of the Observatory from 1 Jan 2004 to 31 Aug 2018.}
\label{fig:effective_areaVsTime}
\end{figure*}

In the case of a steady flux, the exposure ${\cal E}$ to a point-like source of UHE neutrinos multiplied by the spectral flux gives the expected energy distribution of the detected events. The exposure depends on neutrino energy and on the declination $\delta$ of the source and is obtained integrating the effective area ${\cal A}$ over a given time interval:
\begin{equation}
{\cal E}(E_\nu,\delta) = \int_{t_1}^{t_2} {\rm d}t~ {\cal A}(E_\nu,\theta(t),t).
\label{eq:exposure}
\end{equation}
For a given source position $\delta$, the effective area ${\cal A}$ is dependent on $\theta$ (Fig.~\ref{fig:effective_areaVsTheta}) which in turn changes with time as given by Eq.~(\ref{eq:costheta-t}). 
There are also explicit variations of the effective area with time due to the increasing size of the Pierre Auger particle detector array between 2004 and 2008 as the SD stations were being deployed, and due to the exact instantaneous configuration of the array which also varies with time. These latter changes have been relatively small since 2008, the fraction of working stations being typically above $95\%$, as obtained from their continuous monitoring~\cite{Auger_observatory_NIMA2015}. The few periods in which the array has been unstable, mostly due to problems with the communications, are removed from the searches. The time evolution of the effective area (averaged over intervals of three days) is displayed in Fig.~\ref{fig:effective_areaVsTime} during the data-taking period of the Observatory until 31 Aug 2018 for selected values of $\theta$ and energy. The rise during detector deployment is clearly visible before May 2008. The plot also displays a few periods excluded in the following years: 2004, 2005, 2008, 2009, 2014, 2015 and 2016, in which the array was not running stably. Otherwise, the effective area is quite stable after 2008 except for small fluctuations lasting of the order of days.

\begin{figure}[!tb]
\centering
\includegraphics[width=0.9\textwidth]{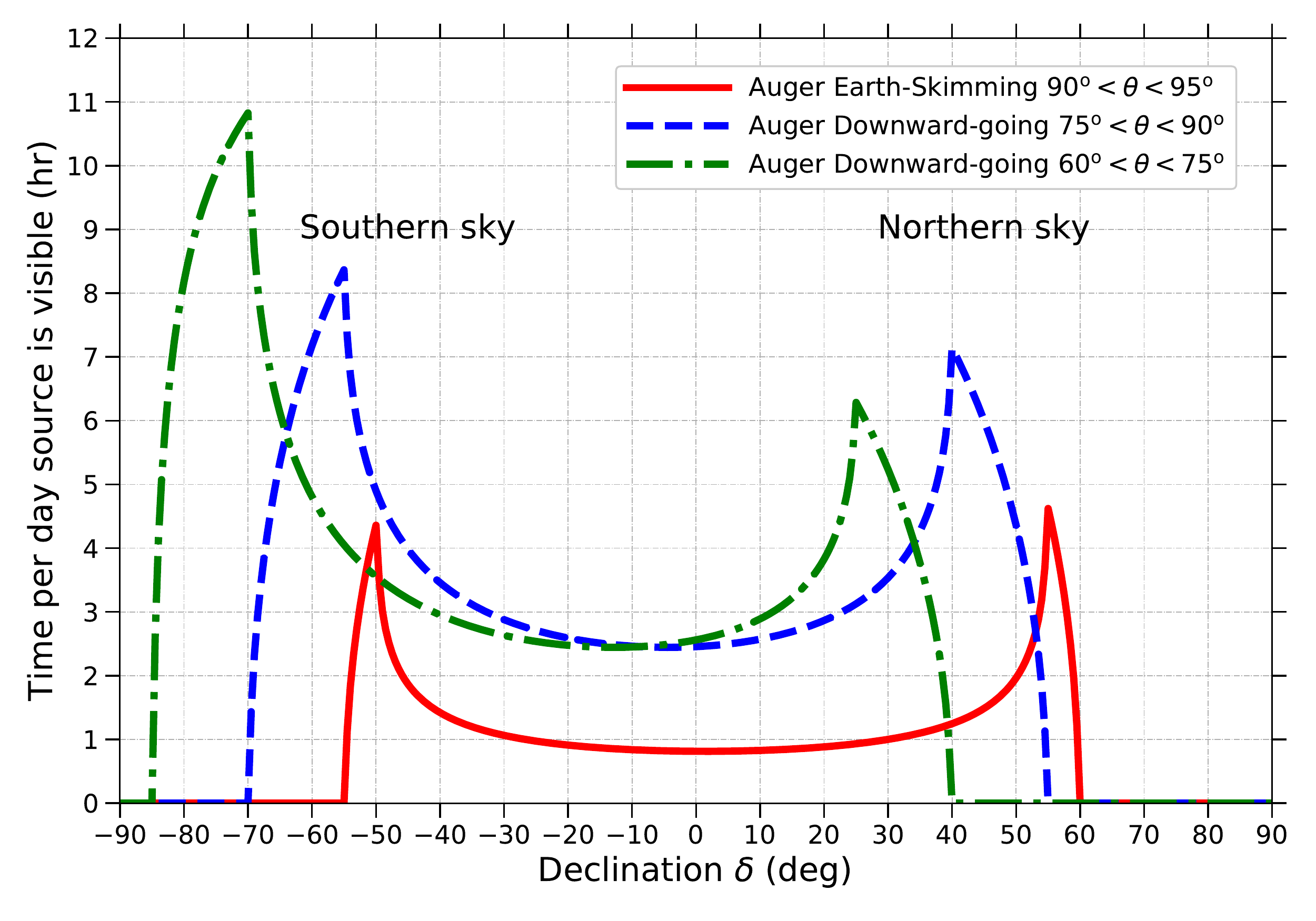}
\caption{Time per sidereal day that a source at declination $\delta$ is visible in the field-of-view of the Earth-Skimming (red line), Downward-going channels (blue line: $75^\circ<\theta<90^\circ$, green line: $60^\circ<\theta<75^\circ$).
}
\label{fig:PS-Fractime}
\end{figure}

When point sources are considered for a neutrino search, the time integral for the exposure in Eq.~(\ref{eq:exposure}) is done with simulations, sampling the array configuration at regular time intervals which are adjusted depending on the total observation period. If the duration of the search interval is of order a day or shorter, the start time of the search interval and its duration take particular relevance for the value of the exposure, due to the changes in effective area with zenith angle, and due to a limited fraction of the sidereal day during which a source at a given declination $\delta$ is within each of the zenith-angle ranges for ES, DGH and DGL neutrino searches. 
Figure~\ref{fig:PS-Fractime} displays the duration of the time interval over which a source is in each of the zenith angle ranges during a sidereal day, obtained with Eq.~(\ref{eq:costheta-t}). 
Near the edges of the field of view of the ES, DGH or DGL channels, there are preferred declination values because the sources are in the field of view for a maximal period of time. These appear as prominent peaks in Fig.~\ref{fig:PS-Fractime}. 
This is a consequence of the relatively slower rate of zenith angle change at these declination values as can be seen in Fig.~\ref{fig:FoVDecHourAngle}. 

\begin{figure}[!tb]
\centering
\includegraphics[width=0.9\textwidth]{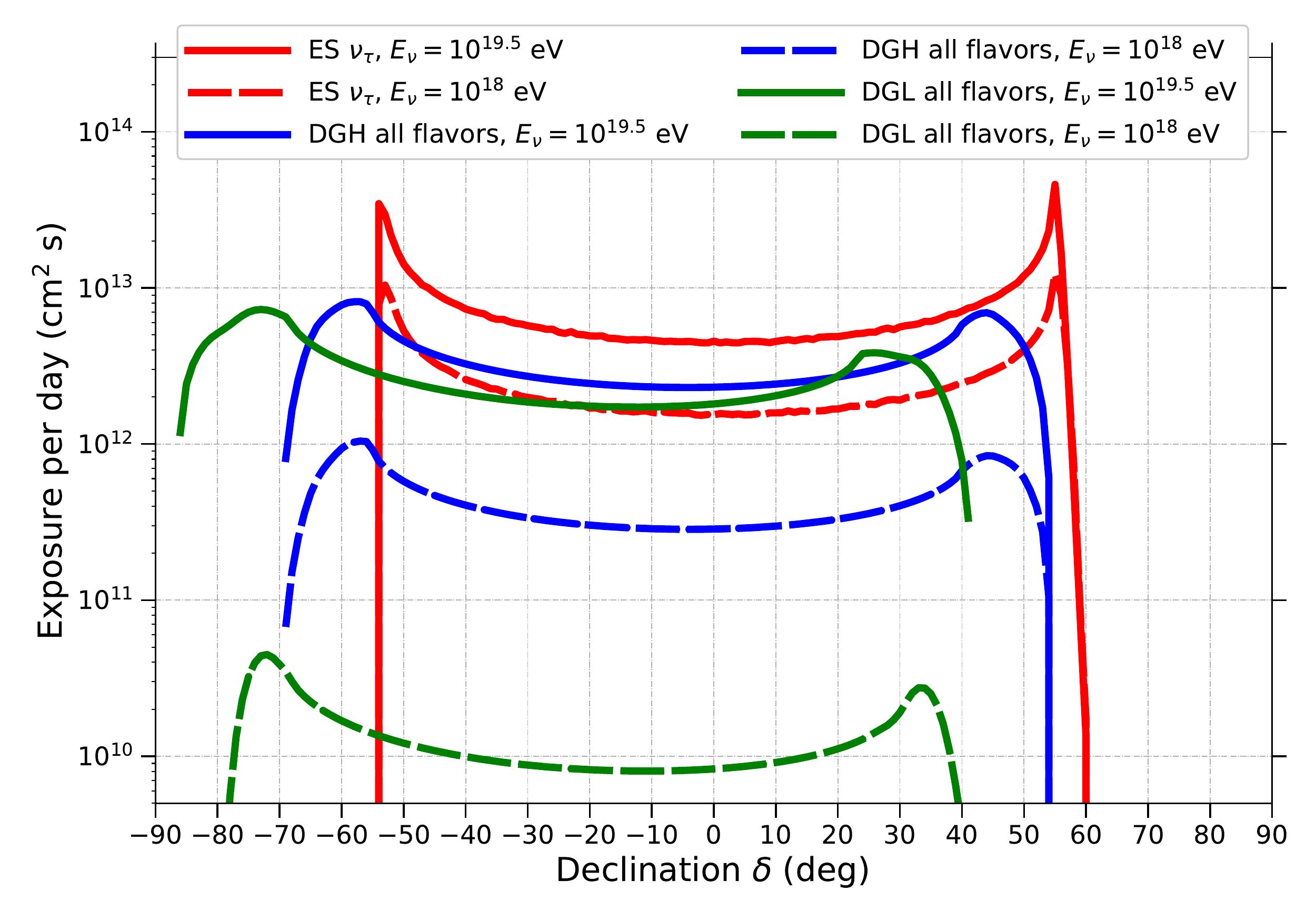}
\caption{Average exposure per day at $E_\nu=10^{18}$~eV and $3\times 10^{19}$~eV as a function of declination $\delta$, calculated for the period from 1 May 2008 when the SD array was completed up to 31 Aug 2018 for the ES, DGH and DGL channels.}
\label{fig:AverageExposure}
\end{figure}

\begin{figure}[!tb]
\centering
\includegraphics[width=0.9\textwidth]{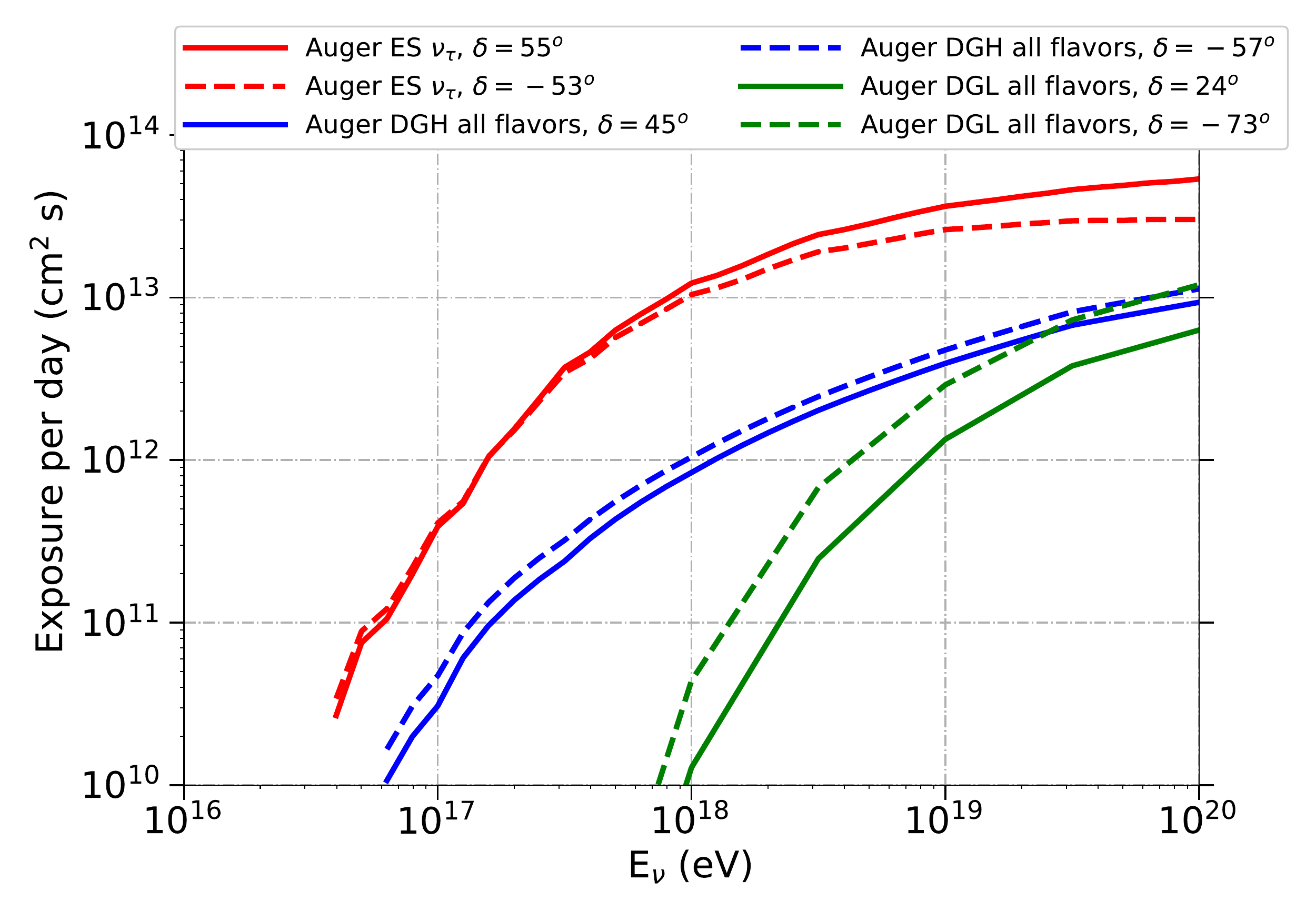}
\caption{Average exposure per day as a function of energy, calculated for the period from 1 May 2008 when the SD array was completed up to 31 Aug 2018 for the pair of declinations at which the exposure in each of the ES, DGH and DGL channels peaks (see Fig.~\ref{fig:AverageExposure}).}
\label{fig:AverageExposure_vs_E}
\end{figure}

The directional exposure obtained daily for the ES, DGH and DGL selections at fixed energies, averaged in the period between 1 May 2008 and 31 August 2018 (excluding the intervals over which the array was unstable) is shown in Fig.~\ref{fig:AverageExposure} as a function of source declination. Complementary, in Fig.~\ref{fig:AverageExposure_vs_E} we show the average exposure as a function of energy for a few selected declinations. When the search interval is much larger than a day, the dependence of the exposure on declination is well approximated by the daily average. The picture shows that there are large variations of the exposure as a function of source declination. Each of the average exposures obtained displays two peaks in declination, close to those that have maximal observation times in Fig.~\ref{fig:PS-Fractime}. In the ES search, the maximal values are obtained for declination values $\delta \sim -53^\circ$ and $\delta \sim 55^\circ$, while in the DGH (DGL) channel the exposure peaks at $\delta \sim -55^\circ$ and $\delta \sim 45^\circ$ ($\delta \sim -70^\circ$ and $\delta \sim 35^\circ$). 
As the effective area for ES neutrinos is larger than that for DGH and DGL (Figs.~\ref{fig:effective_area} and  \ref{fig:effective_areaVsTheta}), the overall largest exposures are obtained for declination values that are close to the peaks in observation time of the ES band alone. 

Two sets of curves for each selection group have been combined in Fig.~\ref{fig:AverageExposure} for $10^{18}$~eV and $10^{19.5}$~eV to illustrate that 
the relative weight of the channels depends strongly on energy. In the EeV range, the effective area for ES neutrinos is larger than that for DGH and DGL, mostly because of the much larger target matter provided by the earth. This translates into average exposure differences of about an order of magnitude for ES relative to DGL despite the transit time per day of a given source being typically shorter for ES. The exposure for DGL is further suppressed by another order of magnitude relative to DGH, mostly because of the efficiency loss for less inclined showers at lower energies. However, the situation is very different at energies of $\sim 3\times 10^{19}$~eV, where the three searches have comparable exposures, and the dominant channel depends on declination. The behaviour of the Pierre Auger Observatory in the search for point sources of neutrinos is thus dependent on the exact energy range of the flux to measure and its spectral features. 

In the case of searches in short time intervals, in addition to this general behaviour of the exposures with declination and energy for each selection group, the position of the source relative to the Observatory at the start point of the search period can play a crucial role. When the source lies just below the horizon, at a zenith angle between $\theta\sim 91^\circ$ and $\theta\sim 93^\circ$ degrees, the effective area is maximal, and the integrated exposure above 0.1 EeV exceeds by an order of magnitude that of IceCube \cite{GW170817_BNS_nus}. 


\section{Limits for steady sources of UHE neutrinos}
\label{sec:results}

The expected number of neutrino events in an energy range $\left[
E_{\rm min},E_{\rm max} \right] $ from a point-like source located at a declination $\delta$ is given by~\footnote{We have here dropped the flavour sub-index $i$; a similar equation can be assumed for each flavour.}:
\begin{equation}
\label{eq:Nevents}
N_{\rm expected}(\delta)=
\int_{E_{\rm min}}^{E_{\rm max}}\!\int_{t_1}^{t_2} 
{\rm d}E_\nu\,{\rm d}t\,  
\phi(E_{\nu},t)\,{\cal A}(E_{\nu},\delta,t).
\end{equation}
To calculate a flux bound for a point source, the spectral function and the time dependence of the neutrino flux of each flavour should be known. In the absence of reliable predictions for these behaviours, it is customary to assume that the flux is independent of time during a given time interval, that the spectral flux has the form of a generic power law $\phi=k_{\rm PS}~E_\nu^{-\alpha}$ with $\alpha=2$, where $k_{\rm PS}$ is the normalization, and that the fluxes of the three neutrino flavours are equal, as expected from vacuum flavour oscillations over scales of hundreds of Mpc~\cite{Learned_APP1995,Athar_PRD2000}. If the flux is independent of time, the exposure integral ${\cal E}(E_\nu,\delta)$, can be factored out and the equation for the number of events simplifies to:
\begin{equation}
\label{eq:Nevents_2}
N_{\rm expected}(\delta)=
\int_{E_{\rm min}}^{E_{\rm max}}{\rm d}E_\nu \, 
\phi(E_{\nu})\,{\cal E}(E_{\nu},\delta).
\end{equation}

For steady fluxes, the time interval is the active period between 1 Jan 2004 and 31 Aug 2018, excluding the unstable periods. 
The lower limit of the energy integral can be taken as zero because the exposure, ${\cal E}$, becomes practically negligible for energies below $\sim 5\times 10^{16}$~eV. Also, under the assumption made for a spectral index of $\alpha = 2$, the bulk of the neutrino triggers is not driven by the upper limit of the energy integral. Most of the identified ES events are between $1.6\times 10^{17}$~eV to $2 \times 10^{19}$~eV while the DG events are between $10^{17}$ and $10^{20}$~eV.
There is little dependence of these energy intervals on source declination or on $E_{\rm max}$ provided that it is larger than $10^{20}$~eV. 

A blind search for UHE neutrinos in the data period up to 31 Aug 2018 has yielded no candidate neutrino events in the ES, DGH, and DGL analyses \cite{Auger_nus_PRD2019}. Under the conservative assumption of zero background, a $90\%$ C.L. upper limit on the neutrino flux from point-like sources is derived assuming $\phi = k_{\rm PS} \cdot E^{-2}_\nu$. The bound on $k_{\rm PS}(\delta)$ is the value that gives a total of 2.39 expected events according to Feldman-Cousins~\cite{Feldman-Cousins} with systematic uncertainties on the exposure calculated using the semi-Bayesian approach described in \cite{Auger_nus_PRD2015}. A bound on $k_{\rm PS}(\delta)$ can be obtained separately for the ES, DGH, and DGL channels. In each of the DG channels, the contributions from different flavours having both NC and CC are combined in the equal flavour assumption. 
In the calculation of the limits, the dependence of the neutrino detection efficiency on the zenith angle and its change with time as the source transits in the field of view of the Pierre Auger Observatory are taken into account. As the data for this work has been taken over the course of multiple years, the exposure can be assumed uniform within $\pm 0.6\%$ in terms of right ascension \cite{Dipole_Science_2017}, and the limits on point-like sources depend solely on the declination. 

\begin{figure*}[!t]
\centering
\includegraphics[width=0.9\textwidth]{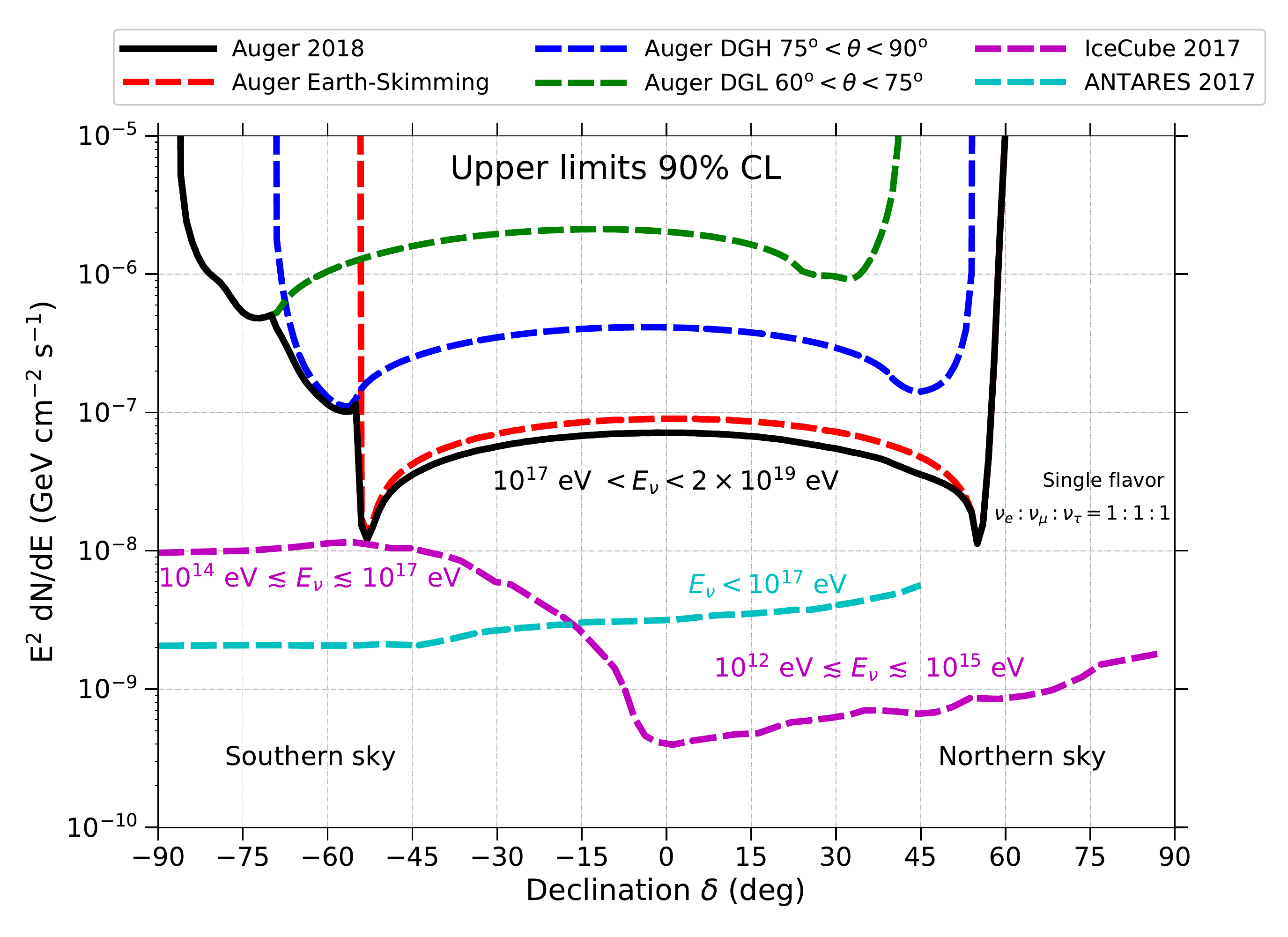}
\caption{Pierre Auger Observatory upper limits (1 Jan 2004 - 31 Aug 2018) at 90$\%$ C.L. on the normalization $k^{\rm PS}$ of
a single flavour point-like flux of UHE neutrinos ${\rm d}N/{\rm d}E_\nu=k^{\rm {PS}} E_\nu^{-2}$ 
as a function of the source declination $\delta$. Also shown are the limits for 
IceCube (2008 - 2015) \cite{IceCube_PS_2017} and ANTARES (2007 - 2015) \cite{ANTARES_PS_2017}. Note the different energy ranges where the limits of each observatory apply.
}
\label{fig:point-like_limits}
\end{figure*}

The limits are shown in Fig.~\ref{fig:point-like_limits} as a function of declination in comparison to those obtained by IceCube~\cite{IceCube_PS_2017} and ANTARES~\cite{ANTARES_PS_2017}. It must be stressed that the energy ranges where the three experiments are sensitive are different and in many respects complementary. The limits reported by ANTARES and IceCube apply to energies just below the energy range of the search for neutrinos with the Pierre Auger Observatory that starts at $\sim10^{17}$~eV. 

\begin{figure}[!t]
\centering
\includegraphics[width=0.9\textwidth]{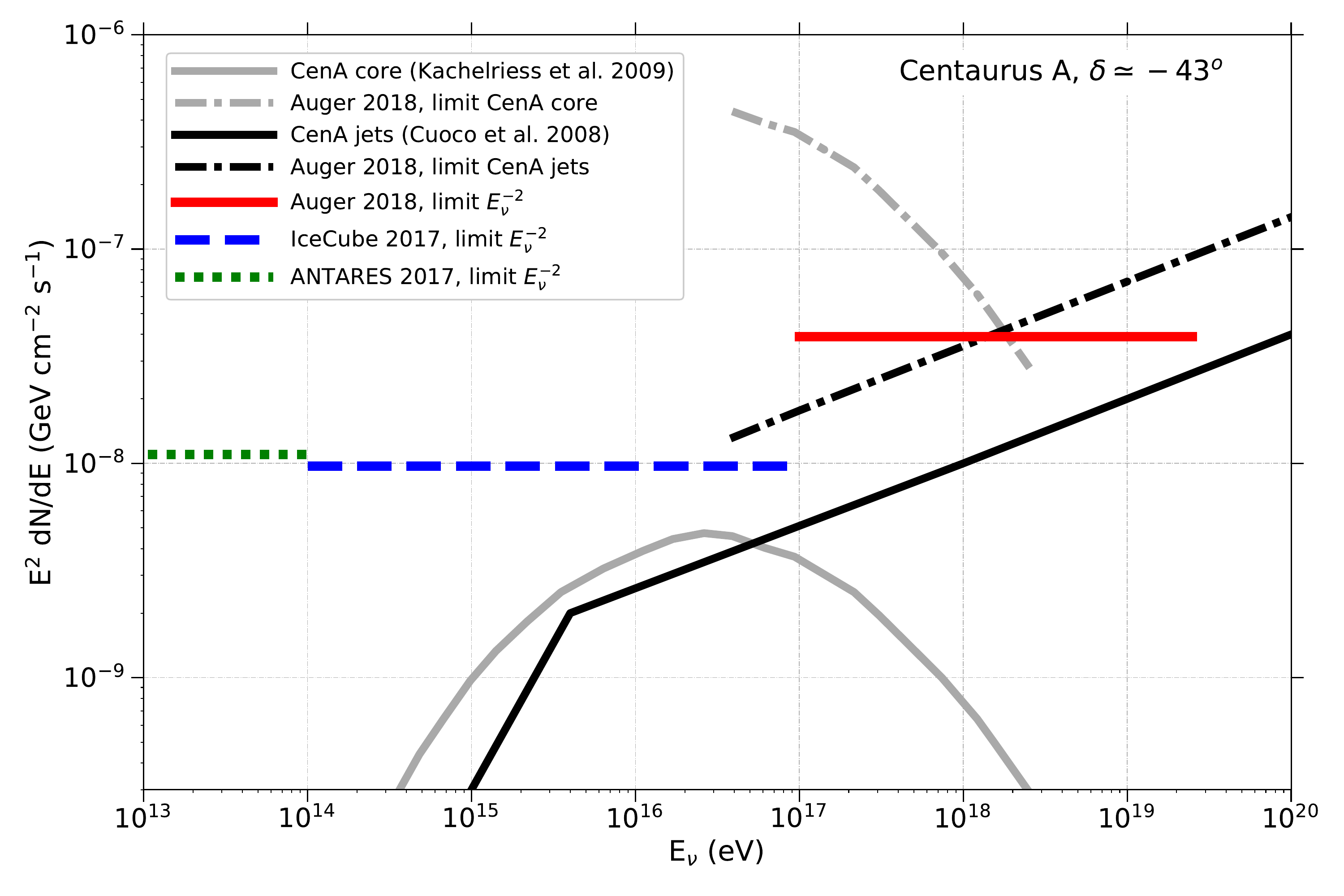}
\vskip -3mm
\caption{
Upper limits at $90\%$ C.L. on a single flavour $E^{-2}$ neutrino flux from the active 
galaxy Centaurus A from the Pierre Auger Observatory, together with limits from IceCube \cite{IceCube_PS_2017} 
and ANTARES \cite{ANTARES_PS_2017}. We also show the predictions of two models of UHE neutrino production in the jets \cite{Cuoco08}, 
and close to the core of Centaurus A \cite{Kachelriess09}, along with $90\%$ C.L. upper limits to these specific models with the 
Auger Observatory.
}
\label{fig:CenA_limits}
\end{figure}

Limits for the particular case of the active galaxy Centaurus A, a potential source of UHECRs, are shown in Fig.~\ref{fig:CenA_limits}, together with constraints from other experiments. CenA at a declination $\delta\sim -43^\circ$ is observed $\sim 7\%$ ($\sim 29\%$) of one sidereal day in the range of zenith angles corresponding to ES (DG) events. The predicted fluxes for two theoretical models of UHE $\nu$-production -- in the jets~\cite{Cuoco08} and close to the core of Centaurus A~\cite{Kachelriess09} -- are also shown. We expect $\sim 0.7$ events from Cen\,A for the flux model in~\cite{Cuoco08} and $\sim 0.025$ events for the model in ~\cite{Kachelriess09}. However, there are significant uncertainties in this model that stem from the fact that the neutrino flux is normalized to the UHECR proton flux assumed to originate from CenA, which is uncertain.

\section{Discussion and Conclusions}
\label{sec:conclusions}

The search for point sources of neutrinos with data from the Surface Detector Array of the Pierre Auger Observatory relies on selecting showers with large zenith angles in three different angular ranges where searches with different sensitivities are performed. The sensitivity of the Observatory to transient sources of UHE neutrinos is demonstrated using the effective area ${\cal A}$ in  Eqs.~(\ref{eq:EffectiveAreaDG}) and (\ref{eq:EffectiveArea_ES}) as an indicator. There are significant differences in effective area for each angular range. The effective area in the Earth-skimming channel has been shown to have a strong dependence on zenith angle, and there is also a strong dependence of the effective areas on neutrino energy. 
This translates into a strong dependence of the exposure for neutrinos on source declination. 

With the Pierre Auger Observatory, we can detect UHE neutrinos from a large fraction of the sky, from very close to the South Celestial Pole to declination values up to $\delta\sim 60^\circ$. For a steady source in this range, there is always a time window during a sidereal day in which the source is in the field of view of the ES, DGH or DGL channels. 

No neutrino candidates have been identified in the Observatory data in the period 1 Jan 2004 to 31 Aug 2018, and 
the limits obtained for steady point-source fluxes represent the most stringent ones at energies around $10^{18}$~eV. The ES channel dominates the sensitivity to point-like sources in the declination region  ($-54.5^\circ  <  \delta < 59.5^\circ$ ). There are two spots in the sky where the sensitivity is maximal, at declinations $\delta \sim -53^\circ$ and $\delta \sim 55^\circ$. The DG channels come into play for smaller declination values making the combined sensitivity cover a large region of the sky
between $\delta \sim -85^\circ$ and $\delta \sim 60^\circ$. 
The SD of the Pierre Auger Observatory has an unmatched sensitivity to potential sources of EeV neutrinos in the Northern terrestrial hemisphere. This is a region in the sky that cannot be searched for in the EeV energy range by experiments such as IceCube because of the opacity of the earth to neutrinos in those directions when seen from the South Pole. 

The sensitivity of the SD of the Pierre Auger Observatory to transient sources of UHE neutrinos is calculated integrating the effective area over the relevant time interval. There is a substantial difference with respect to the search for steady sources of neutrinos. In addition to the declination dependence, the sensitivity to transient sources is crucially dependent on the efficiency of the detection during the time interval of the occurrence of the transient. Thus, depending on the inclination of the event in the local coordinate system of the Observatory, the sensitivity can exceed by far that of other dedicated neutrino detectors such as IceCube. For instance, at about $10^{18}$~eV, the effective area of the Pierre Auger Observatory is maximal for sources slightly below the horizon, and therefore its sensitivity to transients from these directions is larger than that of IceCube by more than an order of magnitude. This is particularly interesting because the location of the GW170817 event, the only confirmed binary neutron star merger to date, was slightly below the horizon at the time around the merger as reported in \cite{GW170817_BNS_nus} and seen in Fig.~\ref{fig:FoVDecHourAngle}. There are other neutrino detectors such as ANITA that actually have a much higher instantaneous effective area~\cite{ANITA_arXiv_2019,
ANITA_PRD_2019}. On the other hand, this detector is limited to flights of about 30 to 40 days during the summer period at Antarctica. 

The Pierre Auger Observatory is in an excellent position to contribute to the new era of multimessenger astronomy, which is likely to bring new exciting discoveries, by looking for neutrinos in the EeV range in correlation with the detection of gamma rays or gravitational waves. 



\section*{Acknowledgments}

\begin{sloppypar}
The successful installation, commissioning, and operation of the Pierre
Auger Observatory would not have been possible without the strong
commitment and effort from the technical and administrative staff in
Malarg\"ue. We are very grateful to the following agencies and
organizations for financial support:
\end{sloppypar}

\begin{sloppypar}
Argentina -- Comisi\'on Nacional de Energ\'\i{}a At\'omica; Agencia Nacional de
Promoci\'on Cient\'\i{}fica y Tecnol\'ogica (ANPCyT); Consejo Nacional de
Investigaciones Cient\'\i{}ficas y T\'ecnicas (CONICET); Gobierno de la
Provincia de Mendoza; Municipalidad de Malarg\"ue; NDM Holdings and Valle
Las Le\~nas; in gratitude for their continuing cooperation over land
access; Australia -- the Australian Research Council; Brazil -- Conselho
Nacional de Desenvolvimento Cient\'\i{}fico e Tecnol\'ogico (CNPq);
Financiadora de Estudos e Projetos (FINEP); Funda\c{c}\~ao de Amparo \`a
Pesquisa do Estado de Rio de Janeiro (FAPERJ); S\~ao Paulo Research
Foundation (FAPESP) Grants No.~2010/07359-6 and No.~1999/05404-3;
Minist\'erio da Ci\^encia, Tecnologia, Inova\c{c}\~oes e Comunica\c{c}\~oes (MCTIC);
Czech Republic -- Grant No.~MSMT CR LTT18004, LO1305, LM2015038 and
CZ.02.1.01/0.0/0.0/16{\textunderscore}013/0001402; France -- Centre de Calcul
IN2P3/CNRS; Centre National de la Recherche Scientifique (CNRS); Conseil
R\'egional Ile-de-France; D\'epartement Physique Nucl\'eaire et Corpusculaire
(PNC-IN2P3/CNRS); D\'epartement Sciences de l'Univers (SDU-INSU/CNRS);
Institut Lagrange de Paris (ILP) Grant No.~LABEX ANR-10-LABX-63 within
the Investissements d'Avenir Programme Grant No.~ANR-11-IDEX-0004-02;
Germany -- Bundesministerium f\"ur Bildung und Forschung (BMBF); Deutsche
Forschungsgemeinschaft (DFG); Finanzministerium Baden-W\"urttemberg;
Helmholtz Alliance for Astroparticle Physics (HAP);
Helmholtz-Gemeinschaft Deutscher Forschungszentren (HGF); Ministerium
f\"ur Innovation, Wissenschaft und Forschung des Landes
Nordrhein-Westfalen; Ministerium f\"ur Wissenschaft, Forschung und Kunst
des Landes Baden-W\"urttemberg; Italy -- Istituto Nazionale di Fisica
Nucleare (INFN); Istituto Nazionale di Astrofisica (INAF); Ministero
dell'Istruzione, dell'Universit\'a e della Ricerca (MIUR); CETEMPS Center
of Excellence; Ministero degli Affari Esteri (MAE); M\'exico -- Consejo
Nacional de Ciencia y Tecnolog\'\i{}a (CONACYT) No.~167733; Universidad
Nacional Aut\'onoma de M\'exico (UNAM); PAPIIT DGAPA-UNAM; The Netherlands
-- Ministry of Education, Culture and Science; Netherlands Organisation
for Scientific Research (NWO); Dutch national e-infrastructure with the
support of SURF Cooperative; Poland -Ministry of Science and Higher
Education, grant No.~DIR/WK/2018/11; National Science Centre, Grants
No.~2013/08/M/ST9/00322, No.~2016/23/B/ST9/01635 and No.~HARMONIA
5--2013/10/M/ST9/00062, UMO-2016/22/M/ST9/00198; Portugal -- Portuguese
national funds and FEDER funds within Programa Operacional Factores de
Competitividade through Funda\c{c}\~ao para a Ci\^encia e a Tecnologia
(COMPETE); Romania -- Romanian Ministry of Research and Innovation
CNCS/CCCDI-UESFISCDI, projects
PN-III-P1-1.2-PCCDI-2017-0839/19PCCDI/2018 and PN18090102 within PNCDI
III; Slovenia -- Slovenian Research Agency, grants P1-0031, P1-0385,
I0-0033, N1-0111; Spain -- Ministerio de Econom\'\i{}a, Industria y
Competitividad (FPA2017-85114-P and FPA2017-85197-P), Xunta de Galicia
(ED431C 2017/07), Junta de Andaluc\'\i{}a (SOMM17/6104/UGR), Feder Funds,
RENATA Red Nacional Tem\'atica de Astropart\'\i{}culas (FPA2015-68783-REDT) and
Mar\'\i{}a de Maeztu Unit of Excellence (MDM-2016-0692); USA -- Department of
Energy, Contracts No.~DE-AC02-07CH11359, No.~DE-FR02-04ER41300,
No.~DE-FG02-99ER41107 and No.~DE-SC0011689; National Science Foundation,
Grant No.~0450696; The Grainger Foundation; Marie Curie-IRSES/EPLANET;
European Particle Physics Latin American Network; and UNESCO.
\end{sloppypar}


\newpage

{\bf\Large{The Pierre Auger Collaboration}}\\
\\

A.~Aab$^{75}$,
P.~Abreu$^{67}$,
M.~Aglietta$^{50,49}$,
I.F.M.~Albuquerque$^{19}$,
J.M.~Albury$^{12}$,
I.~Allekotte$^{1}$,
A.~Almela$^{8,11}$,
J.~Alvarez Castillo$^{63}$,
J.~Alvarez-Mu\~niz$^{74}$,
G.A.~Anastasi$^{42,43}$
% ...


\begin{description}[labelsep=0.2em,align=right,labelwidth=0.7em,labelindent=0em,leftmargin=2em,noitemsep]
\item[$^{1}$] Centro At\'omico Bariloche and Instituto Balseiro (CNEA-UNCuyo-CONICET), San Carlos de Bariloche, Argentina
\item[$^{2}$] Centro de Investigaciones en L\'aseres y Aplicaciones, CITEDEF and CONICET, Villa Martelli, Argentina
\item[$^{3}$] Departamento de F\'\i{}sica and Departamento de Ciencias de la Atm\'osfera y los Oc\'eanos, FCEyN, Universidad de Buenos Aires and CONICET, Buenos Aires, Argentina
\item[$^{4}$] IFLP, Universidad Nacional de La Plata and CONICET, La Plata, Argentina
\item[$^{5}$] Instituto de Astronom\'\i{}a y F\'\i{}sica del Espacio (IAFE, CONICET-UBA), Buenos Aires, Argentina
\item[$^{6}$] Instituto de F\'\i{}sica de Rosario (IFIR) -- CONICET/U.N.R.\ and Facultad de Ciencias Bioqu\'\i{}micas y Farmac\'euticas U.N.R., Rosario, Argentina
\item[$^{7}$] Instituto de Tecnolog\'\i{}as en Detecci\'on y Astropart\'\i{}culas (CNEA, CONICET, UNSAM), and Universidad Tecnol\'ogica Nacional -- Facultad Regional Mendoza (CONICET/CNEA), Mendoza, Argentina
\item[$^{8}$] Instituto de Tecnolog\'\i{}as en Detecci\'on y Astropart\'\i{}culas (CNEA, CONICET, UNSAM), Buenos Aires, Argentina
\item[$^{9}$] Observatorio Pierre Auger, Malarg\"ue, Argentina
\item[$^{10}$] Observatorio Pierre Auger and Comisi\'on Nacional de Energ\'\i{}a At\'omica, Malarg\"ue, Argentina
\item[$^{11}$] Universidad Tecnol\'ogica Nacional -- Facultad Regional Buenos Aires, Buenos Aires, Argentina
\item[$^{12}$] University of Adelaide, Adelaide, S.A., Australia
\item[$^{13}$] Universit\'e Libre de Bruxelles (ULB), Brussels, Belgium
\item[$^{14}$] Vrije Universiteit Brussels, Brussels, Belgium
\item[$^{15}$] Centro Brasileiro de Pesquisas Fisicas, Rio de Janeiro, RJ, Brazil
\item[$^{16}$] Centro Federal de Educa\c{c}\~ao Tecnol\'ogica Celso Suckow da Fonseca, Nova Friburgo, Brazil
\item[$^{17}$] Universidade de S\~ao Paulo, Escola de Engenharia de Lorena, Lorena, SP, Brazil
\item[$^{18}$] Universidade de S\~ao Paulo, Instituto de F\'\i{}sica de S\~ao Carlos, S\~ao Carlos, SP, Brazil
\item[$^{19}$] Universidade de S\~ao Paulo, Instituto de F\'\i{}sica, S\~ao Paulo, SP, Brazil
\item[$^{20}$] Universidade Estadual de Campinas, IFGW, Campinas, SP, Brazil
\item[$^{21}$] Universidade Estadual de Feira de Santana, Feira de Santana, Brazil
\item[$^{22}$] Universidade Federal do ABC, Santo Andr\'e, SP, Brazil
\item[$^{23}$] Universidade Federal do Paran\'a, Setor Palotina, Palotina, Brazil
\item[$^{24}$] Universidade Federal do Rio de Janeiro, Instituto de F\'\i{}sica, Rio de Janeiro, RJ, Brazil
\item[$^{25}$] Universidade Federal do Rio de Janeiro (UFRJ), Observat\'orio do Valongo, Rio de Janeiro, RJ, Brazil
\item[$^{26}$] Universidade Federal Fluminense, EEIMVR, Volta Redonda, RJ, Brazil
\item[$^{27}$] Universidad de Medell\'\i{}n, Medell\'\i{}n, Colombia
\item[$^{28}$] Universidad Industrial de Santander, Bucaramanga, Colombia
\item[$^{29}$] Charles University, Faculty of Mathematics and Physics, Institute of Particle and Nuclear Physics, Prague, Czech Republic
\item[$^{30}$] Institute of Physics of the Czech Academy of Sciences, Prague, Czech Republic
\item[$^{31}$] Palacky University, RCPTM, Olomouc, Czech Republic
\item[$^{32}$] Institut de Physique Nucl\'eaire d'Orsay (IPNO), Universit\'e Paris-Sud, Univ.\ Paris/Saclay, CNRS-IN2P3, Orsay, France
\item[$^{33}$] Laboratoire de Physique Nucl\'eaire et de Hautes Energies (LPNHE), Universit\'es Paris 6 et Paris 7, CNRS-IN2P3, Paris, France
\item[$^{34}$] Univ.\ Grenoble Alpes, CNRS, Grenoble Institute of Engineering Univ.\ Grenoble Alpes, LPSC-IN2P3, 38000 Grenoble, France, France
\item[$^{35}$] Bergische Universit\"at Wuppertal, Department of Physics, Wuppertal, Germany
\item[$^{36}$] Karlsruhe Institute of Technology, Institute for Experimental Particle Physics (ETP), Karlsruhe, Germany
\item[$^{37}$] Karlsruhe Institute of Technology, Institut f\"ur Kernphysik, Karlsruhe, Germany
\item[$^{38}$] Karlsruhe Institute of Technology, Institut f\"ur Prozessdatenverarbeitung und Elektronik, Karlsruhe, Germany
\item[$^{39}$] RWTH Aachen University, III.\ Physikalisches Institut A, Aachen, Germany
\item[$^{40}$] Universit\"at Hamburg, II.\ Institut f\"ur Theoretische Physik, Hamburg, Germany
\item[$^{41}$] Universit\"at Siegen, Fachbereich 7 Physik -- Experimentelle Teilchenphysik, Siegen, Germany
\item[$^{42}$] Gran Sasso Science Institute, L'Aquila, Italy
\item[$^{43}$] INFN Laboratori Nazionali del Gran Sasso, Assergi (L'Aquila), Italy
\item[$^{44}$] INFN, Sezione di Catania, Catania, Italy
\item[$^{45}$] INFN, Sezione di Lecce, Lecce, Italy
\item[$^{46}$] INFN, Sezione di Milano, Milano, Italy
\item[$^{47}$] INFN, Sezione di Napoli, Napoli, Italy
\item[$^{48}$] INFN, Sezione di Roma ``Tor Vergata'', Roma, Italy
\item[$^{49}$] INFN, Sezione di Torino, Torino, Italy
\item[$^{50}$] Osservatorio Astrofisico di Torino (INAF), Torino, Italy
\item[$^{51}$] Politecnico di Milano, Dipartimento di Scienze e Tecnologie Aerospaziali , Milano, Italy
\item[$^{52}$] Universit\`a del Salento, Dipartimento di Matematica e Fisica ``E.\ De Giorgi'', Lecce, Italy
\item[$^{53}$] Universit\`a dell'Aquila, Dipartimento di Scienze Fisiche e Chimiche, L'Aquila, Italy
\item[$^{54}$] Universit\`a di Catania, Dipartimento di Fisica e Astronomia, Catania, Italy
\item[$^{55}$] Universit\`a di Milano, Dipartimento di Fisica, Milano, Italy
\item[$^{56}$] Universit\`a di Napoli ``Federico II'', Dipartimento di Fisica ``Ettore Pancini'', Napoli, Italy
\item[$^{57}$] Universit\`a di Roma ``Tor Vergata'', Dipartimento di Fisica, Roma, Italy
\item[$^{58}$] Universit\`a Torino, Dipartimento di Fisica, Torino, Italy
\item[$^{59}$] Benem\'erita Universidad Aut\'onoma de Puebla, Puebla, M\'exico
\item[$^{60}$] Centro de Investigaci\'on y de Estudios Avanzados del IPN (CINVESTAV), M\'exico, D.F., M\'exico
\item[$^{61}$] Unidad Profesional Interdisciplinaria en Ingenier\'\i{}a y Tecnolog\'\i{}as Avanzadas del Instituto Polit\'ecnico Nacional (UPIITA-IPN), M\'exico, D.F., M\'exico
\item[$^{62}$] Universidad Aut\'onoma de Chiapas, Tuxtla Guti\'errez, Chiapas, M\'exico
\item[$^{63}$] Universidad Nacional Aut\'onoma de M\'exico, M\'exico, D.F., M\'exico
\item[$^{64}$] Institute of Nuclear Physics PAN, Krakow, Poland
\item[$^{65}$] University of \L{}\'od\'z, Faculty of Astrophysics, \L{}\'od\'z, Poland
\item[$^{66}$] University of \L{}\'od\'z, Faculty of High-Energy Astrophysics,\L{}\'od\'z, Poland
\item[$^{67}$] Laborat\'orio de Instrumenta\c{c}\~ao e F\'\i{}sica Experimental de Part\'\i{}culas -- LIP and Instituto Superior T\'ecnico -- IST, Universidade de Lisboa -- UL, Lisboa, Portugal
\item[$^{68}$] ``Horia Hulubei'' National Institute for Physics and Nuclear Engineering, Bucharest-Magurele, Romania
\item[$^{69}$] Institute of Space Science, Bucharest-Magurele, Romania
\item[$^{70}$] University Politehnica of Bucharest, Bucharest, Romania
\item[$^{71}$] Center for Astrophysics and Cosmology (CAC), University of Nova Gorica, Nova Gorica, Slovenia
\item[$^{72}$] Experimental Particle Physics Department, J.\ Stefan Institute, Ljubljana, Slovenia
\item[$^{73}$] Universidad de Granada and C.A.F.P.E., Granada, Spain
\item[$^{74}$] Instituto Galego de F\'\i{}sica de Altas Enerx\'\i{}as (I.G.F.A.E.), Universidad de Santiago de Compostela, Santiago de Compostela, Spain
\item[$^{75}$] IMAPP, Radboud University Nijmegen, Nijmegen, The Netherlands
\item[$^{76}$] KVI -- Center for Advanced Radiation Technology, University of Groningen, Groningen, The Netherlands
\item[$^{77}$] Nationaal Instituut voor Kernfysica en Hoge Energie Fysica (NIKHEF), Science Park, Amsterdam, The Netherlands
\item[$^{78}$] Stichting Astronomisch Onderzoek in Nederland (ASTRON), Dwingeloo, The Netherlands
\item[$^{79}$] Universiteit van Amsterdam, Faculty of Science, Amsterdam, The Netherlands
\item[$^{80}$] Case Western Reserve University, Cleveland, OH, USA
\item[$^{81}$] Colorado School of Mines, Golden, CO, USA
\item[$^{82}$] Department of Physics and Astronomy, Lehman College, City University of New York, Bronx, NY, USA
\item[$^{83}$] Louisiana State University, Baton Rouge, LA, USA
\item[$^{84}$] Michigan Technological University, Houghton, MI, USA
\item[$^{85}$] New York University, New York, NY, USA
\item[$^{86}$] Pennsylvania State University, University Park, PA, USA
\item[$^{87}$] University of Chicago, Enrico Fermi Institute, Chicago, IL, USA
\item[$^{88}$] University of Delaware, Department of Physics and Astronomy, Bartol Research Institute, Newark, DE, USA
\item[$^{89}$] University of Nebraska, Lincoln, NE, USA
\item[] -----
\item[$^{a}$] School of Physics and Astronomy, University of Leeds, Leeds, United Kingdom
\item[$^{b}$] Max-Planck-Institut f\"ur Radioastronomie, Bonn, Germany
\item[$^{c}$] Fermi National Accelerator Laboratory, USA
\item[$^{d}$] also at Universidade Federal de Alfenas, Po\c{c}os de Caldas, Brazil
\item[$^{e}$] Colorado State University, Fort Collins, CO, USA
\item[$^{f}$] now at Hakubi Center for Advanced Research and Graduate School of Science, Kyoto University, Kyoto, Japan
\item[$^{g}$] also at Karlsruhe Institute of Technology, Karlsruhe, Germany
\item[$^{h}$] also at Radboud Universtiy Nijmegen, Nijmegen, The Netherlands
\end{description}

\end{document}